\documentclass[twocolumn]{revtex4}
\usepackage[utf8]{inputenc}
\usepackage[T1]{fontenc}
\usepackage[english]{babel}
\usepackage{amssymb,amsfonts,amsmath}
\usepackage{graphicx}

\begin{document}
\title{Hamiltonian form of extended cubic-quintic nonlinear Schr\"{o}dinger equation \\ in a nonlinear Klein-Gordon model}
\author{Y\lowercase{u}.\,V.~Sedletsky}
\email{sedlets@iop.kiev.ua}
\author{I.\,S.~Gandzha}
\email{gandzha@iop.kiev.ua}
\affiliation{Institute of Physics, National Academy of Sciences of Ukraine, Prosp.~Nauky 46, Kyiv 03028, Ukraine}
\date{\today}

\begin{abstract}
We derive an extended cubic-quintic nonlinear Schr\"{o}dinger equation with Hamiltonian structure in a nonlinear Klein-Gordon model with cubic-quintic nonlinearity. We use the nonlinear dispersion relation to properly take into account the input of high-order nonlinear effects in the Hamiltonian perturbation approach to nonlinear modulation. We demonstrate that changing the balance between the cubic and quintic nonlinearities has a significant effect on the stability of unmodulated wave packets to long-wave modulations.
\end{abstract}

\maketitle

\section{Introduction}

Propagation of nonlinear waves in dispersive media exhibits a variety of fascinating phenomena \cite{Whitham,Ablowitz_Book,Agrawal,Onorato_2016,Bridges_2016}. Modulation instability of plain carrier wave packets is one of such phenomena \cite{Benjamin,Zakharov_2009_MI,Chabchoub_2020_MI}. When dispersion is properly balanced by the nonlinear response of the medium, modulation instability can lead to the formation of envelope solitons representing stable modulated wave packets with localized energy \cite{EnvelopeSoliton}.

When the modulation is slow (wave spectrum is narrow) and wave amplitudes are small as compared to wavelength, the propagation of wave envelopes can be described by the nonlinear Schr\"{o}dinger equation (NLSE) \cite{Sulem_Book,NLSE_Enc,Chabchoub_2015,Copie_2020}. This equation takes into account the second-order dispersion and cubic nonlinearity. As the wave spectrum broadens or wave amplitude grows, high-order dispersion and nonlinear effects start to manifest themselves. Such effects are described by extended (high-order) NLSEs.

There have been two routes of extending the NLSE. The first one is to add the third-order dispersion and the nonlinear dispersion effects described by the first-order derivatives of cubic nonlinearity to get a fourth-order NLSE \cite{Litvak_1967,Dysthe,Kodama_1985,Potasek,Sedlets_NLSE4,Tsitsas_2009,Ferro_2015}. In nonlinear fiber optics, nonlinear dispersion effects are responsible for self-steepening and Raman-induced frequency shift \cite{Agrawal}. The second route is to add the quintic nonlinearity to the cubic NLSE \cite{Pushkarov,Cowan_1986,Malomed_CQ,Christian_2018}, where the emphasis is put on the role of the coupled cubic-quintic nonlinearity, or to the more general fourth-order NLSE \cite{Johnson_Kakutani,Alka_2011,Malomed_2012_Jpn,Chen_2016}. Such models are often referred to as cubic-quintic NLSEs. Sometimes they also include the fourth-order dispersion \cite{Cubic-Quintic_Disp4}.

Cubic-quintic NLSEs represent only a narrow class of more general high-order NLSEs that also take into account higher-order nonlinear dispersion effects described by the second derivatives of cubic nonlinearity \cite{Akhmediev_2014_PRE}. In this context, we refer to such high-order NLSEs as to extended cubic-quintic NLSEs. In particular, extended cubic-quintic NLSEs were considered to describe the dynamics of ferromagnetic spin chains \cite{LPD}, optical solitons \cite{ZakharovKusnetsov}, and water waves \cite{NLSE5_WaterWaves}. NLSE models of yet higher orders have recently been addressed in the literature as well, namely, NLSE with quintic derivative non-Kerr nonlinearities \cite{Choudhuri_2013}, cubic-quintic-septimal NLSE \cite{Cubic-Quintic-Septimal}, sixth-order NLSE \cite{UJP2009,ND18,ND19}, and an hierarchy of integrable high-order NLSEs \cite{NLSE_InfH}.

NLSEs are usually derived by perturbation techniques involving some small physical parameters, most often associated with the smallness of envelope amplitude \cite{Ablowitz_Book}. Such perturbation techniques involve the method of multiple scales \cite{ND18}, the variational method of averaged Lagrangian \cite{ND19}, etc. Introduction of perturbations and small-amplitude expansions usually breaks the Hamiltonian structure of the original Hamiltonian problem, and high-order NLSEs may include non-Hamiltonian terms (although the cubic NLSE is always Hamiltonian) \cite{PRE2020}.

To avoid the origin of non-Hamiltonian terms, one should turn to canonical variables in order to preserve the Hamiltonian structure of governing equations \cite{Zakharov_Hamiltonian}. When applied in Fourier space, Hamiltonian formalism leads to the celebrated four-wave Zakharov integro-differential equation \cite{Zakharov1968} (see also Ref.~\cite{Krasitskii} for more general four- and five-wave forms). NLSE and its high-order extensions can generally be obtained as a narrow-band limit of the Zakharov equations \cite{Gramstad}. Additional conformal mappings and canonical transformations allow the cancellation of certain non-trivial four-wave resonant interactions and produce the so-called compact \cite{Dyachenko_Zakharov_2012} and super compact \cite{Dyachenko_Zakharov_2017} modifications of the Zakharov equations.

Another approach to the Hamiltonian description of nonlinear waves and their modulation was proposed by Craig {\it et al.} \cite{Craig_WM2010} (see also Ref.~\cite{Craig_2021} for its further development). It makes use of the symplectic notation for Hamilton's equations \cite{Goldstein}. The term symplectic means ``intertwined'' and refers to the interlaced role of coordinates and momenta in Hamilton's equations \cite{Procacci_Symplectic}. Accordingly, one can select proper coordinates (called symplectic) that preserve the Hamiltonian character of the original problem \cite{Meyer_Springer}. In this way, Craig {\it et al.} \cite{Craig_WM2010} introduced a complex symplectic coordinate as a coupling of the wave field and momentum in phase space instead of making a transform to normal variables in Fourier space, as is usually done in Zakharov's approach. When written in terms of wave envelopes, such a symplectic coordinate couples the deviation of the wave envelope from equilibrium and a contribution from the motion of envelope with group velocity.

By using an example of a simple physical system described by the nonlinear Klein-Gordon equation with cubic nonlinearity, we have recently demonstrated a relationship between the Hamiltonian form of fourth-order NLSE derived by Craig {\it et al.} in Ref.~\cite{Craig_WM2010} and the non-Hamiltonian form of the same equation \cite{PRE2020}. To this end, we employed the transformation of variables that unambiguously transforms the non-canonical form of fourth-order NLSE for the complex amplitude of the wave field envelope to the canonical form for the envelope of the complex symplectic variable.

The purpose of this work is to add the quintic nonlinearity to the cubic Klein-Gordon model considered in Ref.~\cite{PRE2020} and to extend the Hamiltonian perturbation approach by Craig {\it et al.} \cite{Craig_WM2010} to the case of cubic-quintic nonlinearity. We use the nonlinear dispersion relation to properly take into account the input of high-order nonlinear effects. As a result, we derive the extended cubic-quintic (fifth-order) NLSE in Hamiltonian form that describes the motion of the envelope of coupled wave field and momentum.

From the viewpoint of physics, we are interested in the effect of quintic nonlinearity on the stability of wave packets to long-wave modulations in a Klein-Gordon model with cubic-quintic nonlinearity. When there is no quintic nonlinearity, plain wave packets in such a system are known to be modulationally unstable for any carrier wave number in the case of negative coefficient at cubic nonlinearity. We demonstrate that such plain wave packets become modulationally stable for certain carrier wave numbers when the quintic nonlinearity becomes large enough.

This paper is organized as follows. Section II gives a record of the nonlinear Klein-Gordon model and nonlinear dispersion relation. Section III deals with slow modulation approximation and perturbation expansions. The Hamiltonian form of extended cubic-quintic NLSE is derived in Sect.~IV. Section~V is devoted to the modulation instability condition for the case of cubic-quintic nonlinearity and to the effect of quintic nonlinearity on the stability of uniform wave packets. Conclusions are drawn in Sect.~VI.

\section{Nonlinear Klein-Gordon model and nonlinear dispersion relation}

In this paper we consider a nonlinear Klein-Gordon (nKG) model with cubic-quintic nonlinearity:
\begin{equation}\label{nKG}
\phi_{tt}-c^{2}\phi_{xx}+\alpha_{1}\phi +\alpha_{3}\phi^{3}+\alpha_{5}\phi^{5}=0.
\end{equation}
It can be derived as Hamilton's equations
for the Hamiltonian density
\begin{equation}\label{H}
H=\tfrac{1}{2}\phi_{t}^{2}+\tfrac{1}{2}c^{2}\phi_{x}^{2}+V
\end{equation}
with
$$
V=\tfrac{1}{2}\alpha_{1}\phi^{2}+\tfrac{1}{4}\alpha_{3}\phi^{4}+\tfrac{1}{6}\alpha_{5}\phi^{6}.
$$
Here the unknown real function $\phi$ is a characteristic of the wave field, $t$ is time, $x$ is coordinate, $c$ is the velocity parameter that deals with the speed of interaction propagation. The subscripts denote the partial derivatives. The real coefficient $\alpha_{1}$ describes the linear response of the medium. The real coefficients $\alpha_{3}$ and $\alpha_{5}$ represent the cubic and quintic nonlinearities, respectively.

When $\alpha_{5}=0$, Eq.~(\ref{nKG}) describes the $\phi^4$ model, which is well known in the quantum field theory, elementary particle physics, statistical physics, and condensed matter physics \cite{Rajaraman,Phi4_Book}. The nKG equation with nonzero $\alpha_{5}$ arises in the higher-order $\phi^6$ model \cite{Phi6}. The $\phi^6$ potential possesses three minima (called vacua in the field theory), in contrast to the $\phi^4$ model possessing only two vacua. Field theories of yet higher orders can be formulated as well \cite{Phi8-12}. Finally, when $\alpha_1=1$, $\alpha_3=-\frac{1}{6}$, and $\alpha_5=\frac{1}{120}$, the potential $V$ represents the leading terms of the celebrated sine-Gordon model, which has multiple physical applications \cite{Cuevas,Kevrekidis_2018}.

In the case of weakly nonlinear wave packets, a solution to Eq.~(\ref{nKG}) can approximately be written as a sum of the first and third harmonics:
\begin{equation}\label{phianz}
\phi =\phi_1 +\phi_3 \equiv \varphi +\phi_3,
\end{equation}
with
\begin{align}
\phi_1\equiv\varphi &=\tfrac{1}{2}\Bigl(\varepsilon A\exp\bigl(i(kx-\omega t)\bigr)+\textrm{c.c.}\Bigr),\label{phi1}\\
\phi_{3}&=\tfrac{1}{2}\Bigl(\varepsilon^{3}A_{3}\exp\bigl(3i(kx-\omega t)\bigr)+\textrm{c.c.}\Bigr).  \label{phi3}
\end{align}
Here $k$ and $\omega$ are the wave number and frequency, $\varepsilon$ is a formal small parameter describing the smallness of the wave amplitude, $\varepsilon A$ is the complex amplitude of the first harmonic, $\varepsilon^3 A_3$ is the complex amplitude of the third harmonic, and c.c. denotes the complex conjugate terms. Note that relation (\ref{phianz}) misses the fifth and higher harmonics because they make no contribution to the cubic-quintic NLSE that is a focus of this paper. The zeroth and second harmonics are identically equal to zero when only the odd powers of function $\phi$ are present in the nonlinear part of the nKG equation~(\ref{nKG}).

Substituting function (\ref{phianz}) in Eq.~(\ref{nKG}) yields a nonlinear dispersion relation between the wave frequency and wave number:
\begin{equation}\label{omega2}
\omega^2=c^{2}k^{2}+\alpha_{1}+\tfrac{3}{4}\alpha_{3}\varepsilon^{2}A\overline{A} + O(\varepsilon^4),
\end{equation}
with the bar over $A$ designating the complex conjugate. The well-known linear dispersion relation
\begin{equation}\label{disp_lin}
\omega(k)=\sqrt{c^{2}k^{2}+\alpha_{1}}
\end{equation}
follows as a linear approximation to the more general nonlinear dispersion relation (\ref{omega2}).

Following Craig {\it et al.} \cite{Craig_WM2010}, we introduce the so-called complex symplectic coordinate
\begin{equation}\label{z}
z=\frac{1}{\sqrt{2}}\Bigl(i\frac{1}{\sqrt{\widehat{\omega}}}\,\varphi_{t}+\sqrt{\widehat{\omega}}\varphi \Bigr)
\end{equation}
that is a complex function representing a coupling of the first harmonic $\varphi$ and its derivative $\varphi_t$. The inverse relationship between the functions $\{\varphi,\,\varphi_{t}\}$ and $z$ is given by
\begin{equation}\label{phi}
\varphi =\frac{1}{\sqrt{2\widehat{\omega}}}\left(z+\overline{z}\right),\quad
\varphi_t =\frac{\sqrt{\widehat{\omega}}}{\sqrt{2}i}\left(z-\overline{z}\right).
\end{equation}
Here $\widehat{\omega}$ is a pseudo-differential operator (or the so-called Fourier multiplier operator) such that the wave number $k$ in the dispersion relation is replaced with the differential operator $-i\partial_x$. In the case of linear dispersion relation (\ref{disp_lin}), the operator $\widehat{\omega}$ takes the following form  \cite{Craig_WM2010}:
\begin{equation}\label{omega operator}
\widehat{\omega}=\omega(-i\partial_{x})=\sqrt{c^{2}|{-i\partial_{x}}|^{2}+\alpha_{1}}.
\end{equation}
Note that the term ``pseudo'' refers to the extended nonlocal nature of the operator $\widehat{\omega}$ as compared to ordinary differential operators \cite{Nirenberg_PsiDO,Wong_PsiDO,Lammerzahl_1993}. Roughly speaking, its action on some target function yields a nonpolynomial function of target function itself and its derivative \cite{Fulling_1996}.

Our task is to proceed to the the slow modulation approximation and use the nonlinear dispersion relation instead of the linear one to construct a next-order approximation to the operator $\widehat{\omega}$. The use of the nonlinear dispersion relation is a pivotal step in deriving a consistent fifth-order NLSE as an extension to the fourth-order NLSE derived earlier by Craig {\it et al.} \cite{Craig_WM2010}.

\section{Slow modulation approximation}

Now we proceed with the slow modulation approximation in terms of the complex symplectic coordinate $z$ to derive a Hamiltonian NLSE from the nKG equation (\ref{nKG}). The wave envelope is supposed to be a slow function of time $t$ and coordinate $x$. Therefore, we can introduce the ``slow'' time $\tau =\varepsilon t$ and ``long'' coordinate $\chi =\varepsilon x$ to separate the slow motion of the envelope from fast oscillations of the carrier wave, which are described in terms of the ``fast'' time $t_0 \equiv t$ and ``short'' (normal) coordinate $x_0\equiv x$. Such a mathematical ``trick'' (which is usually referred to as the method of multiple scales) leads to the following perturbation expansions of differential operators:
\begin{equation}\label{dt}
\partial_{t} = \partial_{t_0}+\;\varepsilon \partial_{\tau},\quad
\partial_{x} = \partial_{x_0}+\;\varepsilon \partial_{\chi}.
\end{equation}
Here the formal small parameter $\varepsilon$ is the same as in relations (\ref{phi1}) and (\ref{phi3}) for the functions $\varphi$ and $\phi_3$. With such an approximation, the complex amplitudes $A$ and $A_3$ in (\ref{phi1}) and (\ref{phi3}) are supposed to be slow functions of variables $\chi$ and $\tau$, while the wave phase is supposed to be a fast function of $x_0$ and $t_0$.

The same envelope approximation for the complex symplectic coordinate $z$ is given by
\begin{equation}\label{z_trial}
z=\varepsilon u(\chi,\tau)\exp\bigl(i(k_{0}x_{0}-\omega_{0}t_{0})\bigr),
\end{equation}
where $\varepsilon u$ is the complex amplitude of the envelope of function $z$, $k_0$ is the carrier wave number, and $\omega_0=\omega(k_0)$ is the carrier frequency.

To express the functions $\varphi$ and $\varphi_t$ given by relations (\ref{phi}) in terms of complex amplitude $u$, we need to find a result of action of the operators $\widehat{\omega}^{\frac{1}{2}}$ and $\widehat{\omega}^{-\frac{1}{2}}$ on the complex symplectic coordinate $z$ given by ansatz~(\ref{z_trial}). To this end, we use Theorem~1 from Ref.~\cite{Craig_WM2010} for a Fourier multiplier operator $\widehat{m}$ ($=\widehat{\omega}^{\frac{1}{2}}$ or $\widehat{\omega}^{-\frac{1}{2}}$) and some sufficiently smooth function $f(\chi)$, namely
\begin{multline}\label{Craig}
\widehat{m}(-i\partial_{x})\Bigl(\exp({i k_{0}x})\,f(\chi)\Bigr) \\= \exp({i k_{0}x})\,\widehat{m}\left(k_{0}-i\varepsilon\partial_{\chi}\right)f(\chi).
\end{multline}
This formula basically means the operator expansion around the carrier wave number $k_0$ with the assumption of narrow spectrum and slow modulations.

Next, we expand the operators $\widehat{\omega}^{\frac{1}{2}}$ and $\widehat{\omega}^{-\frac{1}{2}}$ in terms of the formal small parameter $\varepsilon$:
\begin{equation}\label{omega_expand}
\widehat{\omega}^{\pm \frac{1}{2}}\!\left(k_{0}-i\varepsilon\partial_{\chi}\right)
=\omega_0^{\pm \frac{1}{2}}\sum_{n=0}^{\infty}(-i\varepsilon)^{n}\rho^{\pm}_{n}\,\partial_{n\chi},
\end{equation}
where
$$
\rho^{\pm}_{n}=\frac{1}{\omega_0^{\pm \frac{1}{2}}n!}\,\frac{\partial^n}{\partial k^n}\bigl(\omega^{\pm \frac{1}{2}}(k)\bigr)\bigl|_{k=k_{0}}.
$$
The explicit expressions for the first several coefficients $\rho^{\pm}_{n}$ are given in Appendix A.

Operator expansions (\ref{omega_expand}) are calculated with the use of the linear dispersion relation (\ref{disp_lin}). To match these expansions with the nonlinear dispersion relation (\ref{omega2}), we introduce a next-order perturbation to the linear dispersion operator (\ref{omega operator}) as follows:
\begin{multline}\label{operator_omega_nl}
\widehat{\omega}(\varepsilon^2\!A\overline{A})=
\bigl(c^{2}|{-i\partial_{x}}|^{2}+\alpha_{1}+\tfrac{3}{4}\alpha_{3}\varepsilon^{2}A\overline{A}\bigr)^{\frac{1}{2}}\\
\approx\sqrt{c^{2}|{-i\partial_{x}}|^{2}+\alpha_{1}}+\varepsilon^{2}\gamma A\overline{A},\;\; \gamma=\frac{3\alpha_{3}}{8\omega_{0}}.
\end{multline}
Then operator expansions (\ref{omega_expand}) can be extended as
\begin{multline}\label{omega_expand_2}
\widehat{\omega}^{\pm\frac{1}{2}}\!\left(k_{0}-i\varepsilon\partial_{\chi},\,\varepsilon^2 A\overline{A}\right) \\
=\widehat{\omega}^{\pm\frac{1}{2}}\!\left(k_{0}-i\varepsilon\partial_{\chi}\right)\pm
\varepsilon^{2}\omega_0^{\pm \frac{1}{2}}\frac{\gamma}{2\omega_{0}}A\overline{A}.
\end{multline}

Substituting expressions (\ref{phi1}) and (\ref{z_trial}) into relation (\ref{phi}) for $\varphi$ and taking into account formula (\ref{Craig}), the complex amplitude $A$ can be expressed in terms of amplitude $u$:
\begin{equation}\label{Avsu}
A=\sqrt{2}\,\widehat{\omega}^{-\frac{1}{2}}\!\left(k_{0}-i\varepsilon\partial_{\chi}\right)u.
\end{equation}
The linear dispersion approximation to (\ref{Avsu}) can be obtained with operator expansion (\ref{omega_expand}):
\begin{equation*}
A^{(0)}=\sqrt{\frac{2}{\omega_0}}\left(u-i\varepsilon\rho^{-}_{1}u_{\chi}-\varepsilon^{2}\rho^{-}_{2}u_{\chi\chi}+O(\varepsilon^{3})\right).
\end{equation*}
Then the nonlinear dispersion approximation can be derived by substituting the linear dispersion approximation into operator expansion  (\ref{omega_expand_2}):
\begin{multline}\label{A_vs_u}
A=\sqrt{2}\,\widehat{\omega}^{-\frac{1}{2}}\bigl(k_{0}-i\varepsilon\partial_{\chi},\,\varepsilon^2 A^{(0)}\overline{A}\,^{(0)}\bigr)u\\
=\sqrt{\frac{2}{\omega_0}}\Bigl(u-i\varepsilon\rho^{-}_{1}u_{\chi}-\varepsilon^{2}\rho^{-}_{2}u_{\chi\chi}
-\varepsilon^{2}\frac{\gamma}{\omega_0^2}u^2\overline{u}+O(\varepsilon^{3})\Bigr).
\end{multline}
The inverse relationship
\begin{equation}\label{u_vs_A}
u=\sqrt{\frac{\omega _{0}}{2}}\Bigl(A-i\varepsilon\rho^{+}_{1}A_{\chi}-\varepsilon^{2}\rho^{+}_{2}A_{\chi\chi}
+\varepsilon^{2}\frac{\gamma}{2\omega_0}A^2\overline{A}+O(\varepsilon^{3})\Bigr)
\end{equation}
can be derived in a similar way from the relation $z+\overline{z}=\sqrt{2\widehat{\omega}}\,\varphi$, which follows from formula  (\ref{phi}).

Having a relationship between the amplitudes $u$ and $A$, we can proceed straightforward on deriving a high-order NLSE for the amplitude $u$.

\section{Fifth-order NLSE in Hamiltonian form}
The evolution of function $u$ is governed by the equation
\begin{equation}\label{utau}
\varepsilon\omega_{0}u+i\varepsilon^{2}u_{\tau}=\frac{\delta\mathcal{H}}{\delta(\varepsilon\overline{u})}
\end{equation}
that follows from Hamilton's equation for the complex symplectic coordinate $z$, as is demonstrated in Appendix~B. Here
\begin{equation*}
\mathcal{H}=\int \bigl\langle H(u,\overline{u},A_3,\overline{A}_3)\bigr\rangle\,d\chi
\end{equation*}
is the Hamiltonian written in terms of Hamiltonian density (\ref{H}) with the field function $\phi$ represented by ansatz~(\ref{phianz}), $\langle\cdot\rangle$ means averaging over the fast phase $k_{0}x_{0}-\omega_{0}t_{0}$, and $\delta$ denotes the functional derivative.
Due to the symplectic nature of coordinate $z$, the Hamilton equation for the function $\overline{u}$ is just a complex conjugate to the Hamilton equation given by Eq.~(\ref{utau}).

With complex amplitudes $u$ and $\overline{u}$ found from Eq.~(\ref{utau}) and its complex conjugate, variational equations in terms of amplitudes $A_3$ and $\overline{A}_3$ yield the corresponding expressions for these amplitudes in terms of $u$ and $\overline{u}$, namely
\begin{equation}\label{A3u}
A_{3}=\frac{\alpha_3}{32\alpha_1}A^{3}=\frac{\alpha_3}{32\alpha_1}\Bigl(\sqrt{\frac{2}{\omega_{0}}}\,u\Bigr)^{3}+O(\varepsilon).
\end{equation}
The derivation of the above relationship between $A_3$ and $A$ is given in detail in Ref.~\cite{ND19} (see formula (47) therein), and we will not reproduce it here.

To calculate the functional derivative that appears in Eq.~(\ref{utau}), the Hamiltonain density $H$ should be expressed in terms of functions $u$, $\overline{u}$, $A_3$, and $\overline{A}_3$. Below we briefly outline the main steps of how it can be done.

Taking into account ansatz (\ref{phianz}), the Hamiltonian density can be rewritten as a sum of three components, namely,
\begin{equation}\label{Hsum}
H=H_2^{[\varphi]}+H_2^{[\varphi,\phi_3]}+H_{4,6}.
\end{equation}
Here
\begin{equation}\label{H2first_a}
H_2^{[\varphi]}=\tfrac{1}{2}\left(\varphi_{t}^{2}+c^{2}\varphi_{x}^{2}+\alpha_{1}\varphi^{2}\right)
\end{equation}
is the quadratic part of the Hamiltonian density that contains only the first harmonic and
\begin{equation}\label{H2other_a}
H_2^{[\varphi,\phi_3]}=H_{2}-H_2^{[\varphi]}=\tfrac{1}{2}(\phi_{t}^{2}+c^{2}\phi_{x}^{2}+\alpha_{1}\phi^{2})-H_2^{[\varphi]}
\end{equation}
is the remaining part of the quadratic Hamiltonian density that contains both the first and the third harmonics. The non-quadratic part of the Hamiltonian density is designated as
\begin{equation}\label{Hint}
H_{4,6}=H-H_{2}=\tfrac{1}{4}\alpha_{3}\phi^{4}+\tfrac{1}{6}\alpha_{5}\phi ^{6}.
\end{equation}

The quadratic part of the Hamiltonian density given by relation (\ref{H2first_a}) can be expressed in terms of $z$ as follows:
\begin{equation*}
H_2^{[\varphi]} =\tfrac{1}{2}\bigl(i\varphi_t+\widehat{\omega}\,\varphi\bigr)
\bigl(-i\varphi_t +\widehat{\omega}\,\varphi \bigr) =
\sqrt{\widehat{\omega}}z\,\sqrt{\widehat{\omega}}\overline{z}.
\end{equation*}
Taking into account formula (\ref{Craig}), we get an expression
\begin{equation}\label{H2first}
\bigl\langle H_2^{[\varphi]}\bigr\rangle = H_2^{[\varphi]}=
\bigl(\widehat{\omega}^{\frac{1}{2}}\!\left(k_{0}-i\varepsilon\partial_{\chi}\right)u\bigr)
\bigl(\widehat{\omega}^{\frac{1}{2}}\!\left(k_{0}+i\varepsilon\partial_{\chi}\right)\overline{u}\bigr)
\end{equation}
that can easily be expanded using operator expansions (\ref{omega_expand}). Note that here these expansions should be made with the linear dispersion operator because it is the operator that stands in the right-hand side of expression~(\ref{H2first_a}).

The second component of the Hamiltonian density $H_2^{[\varphi,\phi_3]}$ is then calculated by formula (\ref{H2other_a}) with taking into account relation (\ref{H2first}). After some algebraic transformations and averaging over fast phase, we get
\begin{equation}\label{H2_other}
\bigl\langle H_2^{[\varphi,\phi_3]}\bigr\rangle =\tfrac{1}{2}\varepsilon^{6}\left(9c^2k_0^2+5\alpha_{1}\right)A_{3}\overline{A}_{3}+O(\varepsilon^{7}).
\end{equation}
This part of the averaged Hamiltonian density depends only on the amplitude $A_3$ in its leading order of smallness.

The non-quadratic part of the Hamiltonian density yields the following expression after averaging:
\begin{equation}\label{H46}
\bigl\langle H_{4,6}\bigr\rangle=\varepsilon^{4}\bigl\langle H_{4,6}^{(4)}\bigr\rangle+\varepsilon^{5}\bigl\langle H_{4,6}^{(5)}\bigr\rangle+\varepsilon^{6}\bigl\langle H_{4,6}^{(6)}\bigr\rangle+O(\varepsilon^{7}),
\end{equation}
where
\begin{gather*}
\bigl\langle H_{4,6}^{(4)}\bigr\rangle=\frac{3\,\alpha_{3}}{8\,\omega_{0}^{2}}\,u^{2}\overline{u}^{2},\;\;
\bigl\langle H_{4,6}^{(5)}\bigr\rangle=
\frac{3\,\alpha_{3}\omega_{0}^{\prime}}{8\,\omega_{0}^{3}}\left(iu\,\overline{u}^{2}u_{\chi}+\mathrm{c.c.}\right),\\
\bigl\langle H_{4,6}^{(6)}\bigr\rangle=\frac{3\,\alpha_{3}}{32\,\omega_{0}^{4}}\Bigl(
4\,\omega_{0}^{\prime\,2}u\,\overline{u}\,u_{\chi}\overline{u}_{\chi}
-\omega_{0}^{\prime\,2}\left(\overline{u}^{2}u_{\chi}^{2}+\mathrm{c.c.}\right)\\
+\left(2\,\omega_{0}^{\prime\prime}\,\omega_{0}-3\,\omega_{0}^{\prime\,2}\right)\left(u^{2}\overline{u}\,\overline{u}_{\chi\chi}+\mathrm{c.c.}\right)\Bigr)\\
+\frac{5\,\omega_{0}\alpha_{5}-9\,\alpha_{3}\gamma}{12\,\omega_{0}^{4}}u^{3}\overline{u}^{3}
+\frac{\alpha_{3}}{8\,\omega_{0}}\sqrt{\frac{2}{\omega_{0}}}\left(u^{3}\overline{A}_{3}+\mathrm{c.c.}\right).
\end{gather*}
Note that the lowest order of $\bigl\langle H_{4,6}\bigr\rangle$ in (\ref{H46}) is $\varepsilon^4$ because of the fourth power of $\phi$ in expression (\ref{Hint}) for $H_{4,6}$. It is the reason why it was sufficient to keep only the terms of up to order $O(\varepsilon^2)$ in expansion (\ref{A_vs_u}) for $A$ to get the expansion for $\bigl\langle H_{4,6}\bigr\rangle$ up to order $O(\varepsilon^6)$.

Calculating the functional derivative in Eq.~(\ref{utau}) with the averaged Hamiltonian expressed a sum of three components (\ref{H2first}), (\ref{H2_other}), and (\ref{H46}), we finally get a fifth-order NLSE for the complex amplitude $u$:
\begin{equation}\label{NLSE5}
\begin{split}
i\varepsilon^{2}&\bigl(u_{\tau}+\omega'_{0}\,u_{\chi}\bigr)
+\varepsilon^3\bigl(\tfrac{1}{2}\omega''_{0}u_{\chi\chi}+Q^{(3)}|u|^{2}u\bigr)\\
+&\;i\varepsilon^{4}\bigl(-\tfrac{1}{6}\omega'''_{0}u_{\chi\chi\chi}+Q^{(4)}|u|^2u_{\chi}\bigr)\\
+&\;\varepsilon^{5}\bigl(-\tfrac{1}{24}\omega_{0}^{\prime\prime\prime\prime}u_{\chi\chi\chi\chi}
+Q_{1}^{(5)}|u|^{4}u+Q_{2}^{(5)}|u|^2u_{\chi\chi} \\
&\quad+Q_{3}^{(5)}u^{2}\overline{u}_{\chi\chi}+Q_{4}^{(5)}u\,|u_{\chi}|^2+Q_{5}^{(5)}\overline{u}\,u_{\chi}^{2}\bigr)=0,
\end{split}
\end{equation}
where
\begin{align*}
Q^{(3)} &=-\frac{3\,\alpha_{3}}{4\,\omega_{0}^{2}}, \quad
Q^{(4)} = \frac{2\,\omega'_0}{\omega_0}\,Q^{(3)}=\frac{2c^2k}{\omega_0^2}\,Q^{(3)},  \\
Q_{1}^{(5)}&=\frac{9\alpha_{3}\gamma - 5\omega_{0}\alpha_{5}}{4\omega_{0}^{4}}-\frac{3\alpha_3^2}{64\omega_0^3\,\alpha_1},\\
Q_{2}^{(5)}&=\frac{3\alpha_{3}}{4\omega_{0}^{4}}\left(2\omega_{0}^{\prime\,2}-{\omega_{0}^{\prime\prime}}\omega_{0}\right)
=\frac{3\alpha_{3}c^{2}}{4\omega_{0}^{6}}(2c^{2}k_0^{2}-\alpha_{1}),\\
Q_{3}^{(5)}&=\tfrac{1}{2}Q_{4}^{(5)}=\frac{3\alpha_{3}}{8\omega_{0}^{4}}\left(\omega_{0}^{\prime\,2}-{\omega_{0}^{\prime\prime}}\omega_{0}\right)
=\frac{3\alpha_{3}c^{2}}{8\omega_{0}^{6}}(c^{2}k_0^{2}-\alpha_{1}),\\
Q_{5}^{(5)}&=\frac{3\alpha_{3}}{8\omega_{0}^{4}}\left(3\omega_{0}^{\prime\,2}-\omega_{0}^{\prime\prime}\omega_{0}\right)
=\frac{3\alpha_{3}c^{2}}{8\omega_{0}^{6}}(3c^{2}k_0^{2}-\alpha _{1}).
\end{align*}
The coefficients $\omega'_{0}$, $\omega''_{0}$, $\omega'''_{0}$, and $\omega''''_{0}$ are the derivatives of the linear dispersion relation $\omega(k)$ with respect to wave number $k$ calculated at the point $k=k_{0}$ (see Appendix~A). Note that the coefficient $\gamma$ from the nonlinear dispersion operator (\ref{operator_omega_nl}) appears only in the quintic nonlinear coefficient $Q_{1}^{(5)}$. The cubic nonlinear coefficient $Q^{(3)}$ and all the nonlinear derivative coefficients have no contribution from the nonlinear dispersion operator and can correctly be calculated with the linear dispersion operator (\ref{omega operator}).

Note that when referring to a particular order of extended NLSE we mean the aggregate order of the wave amplitude and spatial derivative entering the highest order of the equation. Such a classification is well established in the field of nonlinear water waves \cite{Gramstad}. Within this classification, the classical cubic NLSE is referred to as the third-order NLSE. The equation with the third-order dispersion and first-order cubic derivative terms is referred to as the fourth-order NLSE. In particular, the celebrated Dysthe equation \cite{Dysthe} describing the modulations of deep-water waves is well known as the fourth-order NLSE \cite{Shemer_Dysthe}. Finally, the equation with the fourth-order dispersion and quintic nonlinearity is referred to as the fifth-order NLSE. This remark is made to avoid any misunderstanding because the reader can meet an alternative classification in the literature (see, e.g., Ref.~\cite{NLSE_InfH}) where high-order NLSEs are classified by the order of the highest dispersion term entering the equation.

From the technical point of view, the extended cubic-quintic NLSE (\ref{NLSE5}) with its coefficients expressed in terms of parameters of the original nKG equation is the main result of this paper. As compared to the extended NLSE considered in Ref.~\cite{PRE2020}, Eq.~(\ref{NLSE5}) contains six additional terms that describe the fourth-order dispersion, quintic nonlinearity, and second-order cubic nonlinear dispersion effects. Equation (\ref{NLSE5}) is a Hamiltonian PDE inasmuch as it was derived from Hamilton's equation. Its second-order cubic nonlinear derivative coefficients satisfy the condition
\begin{equation}
Q_{2}^{(5)}-2Q_{3}^{(5)}+\tfrac{1}{2}Q_{4}^{(5)}=Q_{5}^{(5)}
\end{equation}
that should hold for Eq.~(\ref{NLSE5}) to have a Hamiltonian. To the best of our knowledge, the fifth-order NLSE in Hamiltonian form~(\ref{NLSE5}) has not previously been reported in the literature in the context of the nKG model.

The Hamiltonian nature of Eq.~(\ref{NLSE5}) is due to the use of symplectic coordinate $z$ that couples the wave field $\phi$ and its momentum $p=\phi_t$. In this case the symplectic coordinate $z$ and its complex conjugate $\overline{z}$ form a pair of complex canonical variables. In quantum mechanics, the use of such variables corresponds to a transition from the coordinate-momentum representation to a representation involving the creation and annihilation Bose operators. In the framework of slow modulation approximation, the complex amplitude $u$ of the envelope of canonical variable $z$ couples the deviation of the wave field envelope from equilibrium and a contribution from the motion of envelope with group velocity $v_g=\omega_0'$, namely
\begin{equation*}
\sqrt{\frac{2}{\omega_0}}u= A-\varepsilon\frac{iv_g}{2\omega_0}A_{\chi}+\ldots,
\end{equation*}
as follows from relation (\ref{u_vs_A}). In contrast to the Hamiltonian equation (\ref{NLSE5}) for $u$, the evolution equation for the uncoupled (i.e., non-canonical) amplitude $A$ is non-Hamiltonian, as is shown in Appendix~C. Therefore, the above coupling is a pivotal step in deriving a high-order NLSE in Hamiltonian form.

The coefficients at cubic, quintic, and derivative nonlinear terms of Eq.~(\ref{NLSE5}) all depend on the cubic coefficient $\alpha_3$ of the nKG equation. On the other hand, the quintic coefficient $\alpha_5$ enters Eq.~(\ref{NLSE5}) only through the coefficient $Q_{1}^{(5)}$ of the quintic nonlinear term. A natural question that arises in this context is whether there is any significant effect of including the quintic nonlinearity in the nKG model and at which conditions it can manifest itself. We address this question in the next section and demonstrate that the quintic coefficient $\alpha_5$ has a significant effect on the modulation instability of a homogeneous (constant-amplitude) solution to Eq.~(\ref{NLSE5}).

\section{Modulation instability}
In this section we demonstrate that the quintic coefficient $\alpha_5$ significantly modifies the modulation instability condition and results in the formation of stability regions absent in the case when only the cubic coefficient $\alpha_3$ is considered. Assuming that $\alpha_1>0$, we introduce the dimensionless time and coordinate
\begin{equation}\label{TX}
T = \sqrt{\alpha_1}\,t =\varepsilon^{-1}\sqrt{\alpha_1}\,\tau, \quad
X = \varepsilon^{-1}\frac{\sqrt{\alpha_1}}{c}\,\chi
\end{equation}
and put Eq.~(\ref{NLSE5}) in dimensionless form for the rescaled amplitude $\psi = \varepsilon u$:
\begin{equation}\label{NLSE5_psi}
\begin{split}
&i\bigl(\psi_{T}+\beta_1\,\psi_{X}\bigr)+\beta_2\,\psi_{X\!X}+\mathcal{Q}^{(3)}|\psi|^{2}\psi\\
&+i\bigl(-\beta_3\,\psi_{X\!X\!X}+\mathcal{Q}^{(4)}|\psi|^2\psi_{X}\bigr)\\
&-\beta_4\,\psi_{X\!X\!X\!X}+\mathcal{Q}_{1}^{(5)}|\psi|^{4}\psi+\mathcal{Q}_{2}^{(5)}|\psi|^2\psi_{X\!X} \\
&\quad+\mathcal{Q}_{3}^{(5)}\psi^{2}\overline{\psi}_{X\!X}+\mathcal{Q}_{4}^{(5)}\psi\,|\psi_{X}|^2+\mathcal{Q}_{5}^{(5)}\overline{\psi}\,\psi_{X}^{2}=0.
\end{split}
\end{equation}
The rescaled nonlinear coefficients of this equation are expressed as
\begin{align*}
\mathcal{Q}^{(3)} &=\frac{1}{\sqrt{\alpha_1}}\,Q^{(3)}=-\frac{3\,\tilde\alpha_{3}}{4\,\tilde\omega^{2}}, \quad \mathcal{Q}^{(4)} = \frac{2\kappa}{\tilde\omega^2}\,\mathcal{Q}^{(3)},  \\
\mathcal{Q}_{1}^{(5)}&=\frac{1}{\sqrt{\alpha_1}}\,Q_1^{(5)}=\frac{51-80\,\tilde\omega^2\tilde\alpha_5-3\kappa^2}{64\,\tilde\omega^5}\;\tilde\alpha_3^2,\\
\mathcal{Q}_{2}^{(5)}&=\frac{\sqrt{\alpha_1}}{c^2}\,Q_2^{(5)}=\frac{1-2\kappa^2}{\tilde\omega^{4}}\;\mathcal{Q}^{(3)},\\
\mathcal{Q}_{3}^{(5)}&=\tfrac{1}{2}\mathcal{Q}_{4}^{(5)}=\frac{\sqrt{\alpha_1}}{c^2}\,Q_3^{(5)}=\frac{1-\kappa^2}{2\,\tilde\omega^{4}}\;\mathcal{Q}^{(3)},\\
\mathcal{Q}_{5}^{(5)}&=\frac{\sqrt{\alpha_1}}{c^2}\,Q_5^{(5)}=\frac{1-3\kappa^2}{2\,\tilde\omega^{4}}\;\mathcal{Q}^{(3)},
\end{align*}
where
\begin{equation}\label{kappa}
\tilde{\omega}=\frac{\omega_0}{\sqrt{\alpha_1}}=\sqrt{\kappa^2+1},\quad\kappa=\frac{ck_0}{\sqrt{\alpha_1}}
\end{equation}
are the dimensionless carrier frequency and wave number. The coefficients $\beta_n = \frac{1}{n!}\frac{\partial^n\tilde\omega}{\partial\kappa^n}$ account for the linear dispersion contribution. The dimensionless parameters
\begin{equation}\label{tildealpha}
\tilde\alpha_{3}=\frac{\alpha_{3}}{\sqrt{\alpha_{1}^{3}}},\quad \tilde\alpha_{5}=\frac{\alpha_{5}\alpha_1}{\alpha_{3}^{2}}
\end{equation}
are the scaled cubic and quintic coefficients of the nKG equation (\ref{nKG}) (here we assume that $\alpha_3\ne0$).

\begin{figure}[!]
\includegraphics[width=0.8\columnwidth]{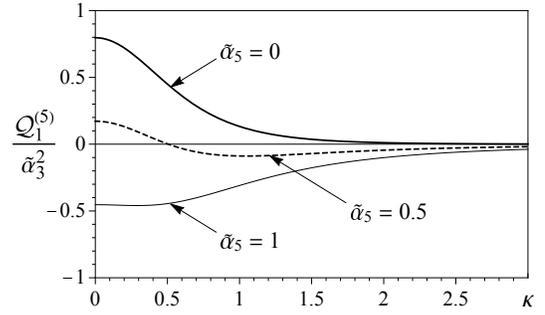}
\caption{\label{fig1} Scaled coefficient $\mathcal{Q}_{1}^{(5)}$ of the extended NLSE (\ref{NLSE5_psi}) as a function of dimensionless carrier wave number $\kappa$ for three different values of parameter $\tilde\alpha_{5}$.}
\end{figure}

The coefficient $\mathcal{Q}_{1}^{(5)}$ is the only coefficient of Eq.~(\ref{NLSE5_psi}) that depends on quintic coefficient $\alpha_{5}$. Figure~\ref{fig1} shows the scaled coefficient $\mathcal{Q}_{1}^{(5)}$ as a function of dimensionless wave number $\kappa$ for three different values of parameter $\tilde\alpha_{5}$. When $\tilde\alpha_{5}=0$ (no quintic nonlinearity), the coefficient $\mathcal{Q}_{1}^{(5)}$ stays positive for any $\kappa$. As $\tilde\alpha_{5}$ increases from zero, the coefficient $\mathcal{Q}_{1}^{(5)}$ changes its sign at some nonzero $\kappa$. Finally, it becomes negative for any $\kappa$ when $\tilde\alpha_{5}$ approaches unity. Thus, the quintic coefficient $\alpha_5$ of the nKG equation (\ref{nKG}) has a profound effect on the quintic nonlinear coefficient of the extended cubic-quintic NLSE~(\ref{NLSE5_psi}) when $\alpha_5$ is nearly as large as the ratio $\alpha_3^2/\alpha_1$ or is exceeding it. Hereafter we elaborate upon this finding and demonstrate the effect of coefficient $\alpha_5$ on the modulation instability of a homogeneous (constant-amplitude) solution to Eq.~(\ref{NLSE5_psi}).

\begin{figure*}[!]
\includegraphics[width=\textwidth]{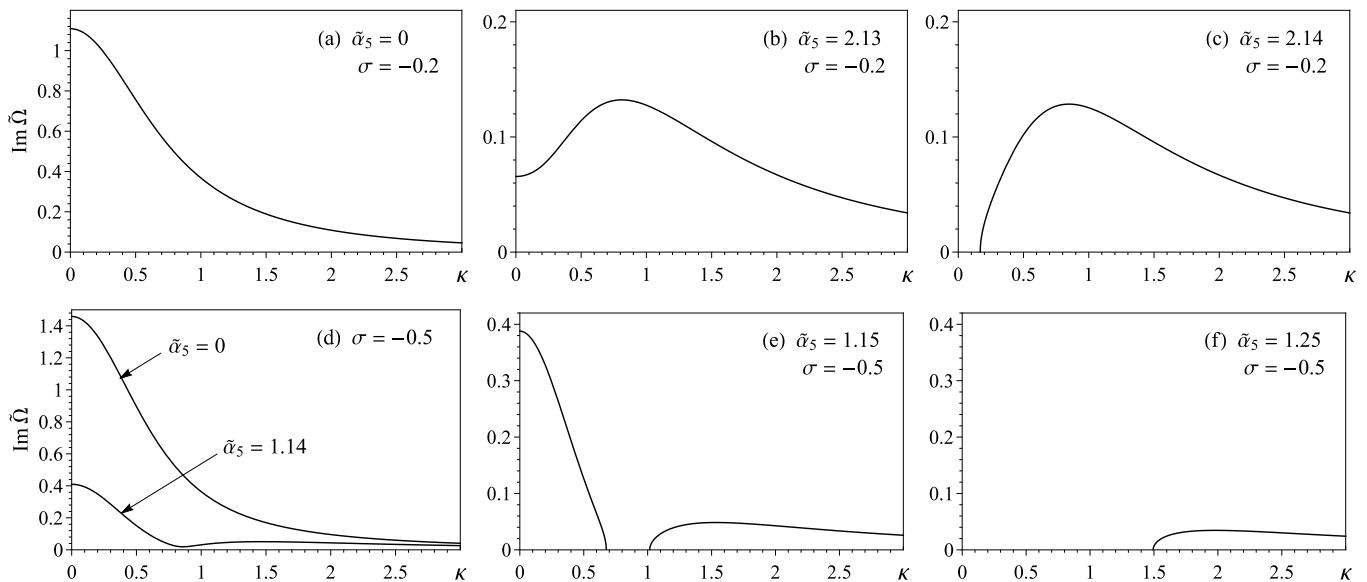}
\caption{\label{fig2}Scaled initial growth rate $\mathrm{Im}\,\tilde\Omega$ of long-wave modulations of uniform wave packets as a function of dimensionless carrier wave number $\kappa$ for two values of parameter $\sigma=\tilde\alpha_3|\psi_0|^2$: (a,b,c) $\sigma=-0.2$ and (d,e,f) $\sigma=-0.5$. Constant-amplitude carrier wave packets are modulationally unstable ($\mathrm{Im}\,\tilde\Omega>0$) for any $\kappa$ when there is no quintic nonlinearity ($\tilde\alpha_5 = 0$). Stability regions ($\mathrm{Im}\,\tilde\Omega=0$) are formed at some critical value of $\tilde\alpha_5$ that depends on $\sigma$: $\tilde\alpha_5^{(\mathrm{c})}\approx2.138$ for $\sigma=-0.2$ and $\tilde\alpha_5^{(\mathrm{c})}\approx1.142$ for $\sigma=-0.5$.}
\end{figure*}

Modulation instability means the instability of a constant-amplitude wave packet to long-wave modulations. The unmodulated solution to Eq.~(\ref{NLSE5_psi}) is given by a homogeneous constant-amplitude wave function
\begin{equation}\label{eq:homogen}
\psi(T)=\psi_0\exp\bigl(i\mu\,|\psi_0|^2 T\bigr),
\end{equation}
where $\mu=\mathcal{Q}^{(3)} + \mathcal{Q}_1^{(5)}|\psi_0|^2$.
The condition of modulation instability of homogeneous solution (\ref{eq:homogen}) can be determined by introducing a small
perturbation to the complex amplitude $\psi_0$:
\begin{equation}
\psi(X,T)=\bigl(\psi_0+\epsilon(X,T)\bigr)\exp\bigl(i\mu\,|\psi_0|^2 T\bigr),
\end{equation}
where
\begin{equation*}
\epsilon(X,T)=\epsilon_0^{+}\exp\left(i\varkappa X-i\,\Omega\,T\right)+
\epsilon_0^{-}\exp\left(i\,\overline{\Omega}\,T-i\varkappa X\right).
\end{equation*}
Here we assume the perturbation frequency $\Omega$ to be
complex-valued and the perturbation wave number $\varkappa$ to be real.
Substituting this ansatz in Eq.~(\ref{NLSE5_psi}) leads to the
following relationship between $\Omega$ and $\varkappa$:
\begin{equation}\label{eq:MI_Omega}
\Omega = \bigl(\beta_1+\mathcal{Q}^{(4)}|\psi_0|^2\bigr)\varkappa + \beta_3\varkappa^3+\beta_4\varkappa^4\pm |\varkappa|\sqrt{S},
\end{equation}
where $S=s_1+s_2\varkappa^2$ and
\begin{align*}
s_1&=-2\bigl(\mathcal{Q}^{(3)}\!\! + 2\mathcal{Q}_1^{(5)}|\psi_0|^2\bigr)\bigl(\beta_2\!+
(\mathcal{Q}_2^{(5)}\!\!-\!\mathcal{Q}_3^{(5)})|\psi_0|^2\bigr)|\psi_0|^2,\\
s_2&=\beta_2^2+2\beta_2\mathcal{Q}_2^{(5)}|\psi_0|^2+\bigl(\bigl(\mathcal{Q}_2^{(5)}\bigr)^2\!-\bigl(\mathcal{Q}_3^{(5)}\bigr)^2\bigr)\,|\psi_0|^4.
\end{align*}
A homogeneous solution is modulationally unstable when $\mathrm{Im}\,\Omega>0$ (perturbation exponentially grows with time). Since the first three terms in formula~(\ref{eq:MI_Omega}) for $\Omega$ are real, the condition of modulation istability effectively requires the radicand $S$ to be negative.

Considering only long-wave modulations, we can require that $\varkappa\rightarrow0$. In this case, the radicand $S$ can be expressed in the following explicit form:
\begin{multline}
S=s_1=\frac{\sigma}{128\,\tilde\omega^{11}}\bigl((80\,\tilde\omega^2\tilde\alpha_5+3\kappa^2-51)\,\sigma+24\,\tilde\omega^3\bigr)\times\\
\bigl((3\kappa^2-1)\,3\sigma+4\,\tilde\omega^3\bigr),\quad \sigma=\tilde\alpha_3\,|\psi_0|^2.
\end{multline}
Then the initial growth rate of modulations can be calculated as
\begin{equation}
\mathrm{Im}\,\Omega = |\varkappa||\sigma|^{\frac{1}{2}}\,\mathrm{Im}\,\tilde\Omega,\quad
\tilde\Omega = \sqrt{|\sigma|^{-1}s_1}.
\end{equation}
The scaled frequency $\tilde\Omega$ is a function of three dimensionless parameters: carrier wave number $\kappa$, scaled quintic coefficient $\tilde\alpha_5$, and parameter $\sigma$.

The dimensionless parameter $\sigma$ is proportional to the squared absolute value of the wave function's amplitude and to the cubic coefficient $\alpha_3$ (since $\tilde\alpha_3\propto\alpha_3$). It is positive when $\alpha_3>0$ and negative when $\alpha_3<0$. To determine the possible range of parameter $\sigma$, we need to recall that the amplitudes of the first and third harmonics in the approximate solution (\ref{phianz}) to the nKG equation were assumed to be small. This condition implies that the ratio between the amplitudes of the third and first harmonics needs to be small:
\begin{equation*}
\frac{|\phi_3|}{|\phi_1|}=\varepsilon^2\left|\frac{A_3}{A}\right|\ll 1.
\end{equation*}
Taking into account relation (\ref{A3u}) for $A_3$, we come to the following condition that should be imposed on the parameter $\sigma$:
$|\sigma|\ll 16\,\tilde\omega$. From the practical point of view, this condition holds for any $|\sigma|\lesssim1$.

Here we restrict our consideration to the case $\sigma<0$ (negative $\alpha_3$) and consider two different values of parameter $\sigma$, namely, $\sigma=-0.2$ and $\sigma=-0.5$. The case of positive $\sigma$ can be considered in a similar manner. Figure~\ref{fig2} shows the scaled initial growth rate $\mathrm{Im}\,\tilde\Omega$ of long-wave modulations as a function of dimensionless carrier wave number $\kappa$ for the two values of $\sigma$ under consideration. When $\tilde\alpha_5=0$ (no quintic nonlinearity), the constant-amplitude solution (\ref{eq:homogen}) is modulationally unstable ($\mathrm{Im}\,\tilde\Omega>0$) at any $\kappa$ for both $\sigma=-0.2$ (Fig.~\ref{fig2}(a)) and $\sigma=-0.5$ (Fig.~\ref{fig2}(d)). When the quintic nonlinearity becomes large enough, there appears the stability region ($\mathrm{Im}\,\tilde\Omega=0$) that is absent in the case of dominant cubic nonlinearity. When $\sigma=-0.2$ (Fig.~\ref{fig2}(b,c)), this stability region is formed for long waves ($\kappa\rightarrow0$) and then enlarges in the direction of shorter waves (larger $\kappa$). In the case of $\sigma=-0.5$, the stability region is first formed in the vicinity of $\kappa\approx0.85$ (Fig.~\ref{fig2}(d)) and then enlarges both in the direction of shorter and longer waves (Fig.~\ref{fig2}(e)), until all long waves (small $\kappa$) become modulationally stable (Fig.~\ref{fig2}(f)).

These results demonstrate that the quintic nonlinearity, when it is large enough, dramatically changes the stability of uniform carrier wave packets to long-wave modulations. 

\section{Conclusions}

We considered a Klein-Gordon model with cubic-quintic nonlinearity that describes a relativistic scalar field with a quartic-sextic potential. From the viewpoint of elementary particle physics, this model represents a relativistic field equation for spinless scalar particles in a quartic-sextic potential. The stationary version of this equation can also be used to describe the macroscopic wave function of the condensed phase (i.e., the order parameter) in the Ginzburg-Landau theory of superconductivity. In this case, the potential (quartic, sextic, or even of higher order) is interpreted as the Landau free energy density, with its minima (equilibrium positions) defining the parent (high temperature) and product (low temperature) phases \cite{Sanati_Saxena_1999}. Structural changes in the form of the potential under variations in the control parameter (which is usually associated with temperature) describe different types of phase transitions in such a system. A spatial gradient of the order parameter (Ginzburg term) allows for the existence of domain walls (or the so-called kinks) between various phases (see Chap.~12 of Ref.~\cite{Phi4_Book} for more details).

Unlike the above-mentioned studies that mainly address the stationary regimes of the nonlinear Klein-Gordon (nKG) model, we are interested in nonstationary effects arising from the evolution of the wave field in time. Here we have studied the envelope properties of the nKG model by transforming it into a high-order (extended cubic-quintic) NLSE model. A remarkable feature of this work is that the high-order NLSE was obtained in Hamiltonian form, as opposed to the previous works \cite{ND18,ND19} on this subject. To this end, we extended the Hamiltonian perturbation approach to nonlinear modulation proposed by Craig {\it et al.} \cite{Craig_WM2010} and used the nonlinear dispersion relation to properly take into account the input of high-order nonlinear effects. Drawing an analogy with quantum mechanics, our approach corresponds to a transition from the coordinate-momentum representation to a representation involving the creation and annihilation Bose operators.

When the carrier wave packets are modulationally unstable, NLSE is known to admit envelope solitons (or quasi-solitons in high-order NLSE models). These localized wave structures can be interpreted as bound states of quasiparticles represented by plane waves \cite{Sharma_Buti}.
Here we demonstrated that the quintic nonlinearity of the nKG model significantly modifies the modulation instability condition and results in the formation of stability regions absent in the case when only the cubic nonlinearity is considered. This happens for certain wave numbers at a certain threshold in the ratio of the quintic and cubic coefficients of the nKG equation. Thus, the existence conditions for envelope solitons in a system with quintic nonlinearity may break for those wave numbers where the carrier waves are modulationally stable.

We believe these results will facilitate further studies of nonstationary phenomena in physical systems involving cubic-quintic nonlinearities and quartic-sextic potentials.

\acknowledgments{
This manuscript was prepared during the ongoing war in Ukraine. We are deeply thankful to all the brave people who have been fighting for the freedom and independence of our nation and country.}

\appendix

\section{Coefficients of expansions}

Several leading coefficients of operator expansions (\ref{omega_expand}) are as follows:
\begin{align*}
\rho^+_{0}&=\rho^-_{0}=1,\quad
\rho^+_{1}=-\rho^-_{1}=\frac{\omega_{0}^{\prime}}{2\omega_{0}},\\
\rho^+_{2}&=\frac{\omega_{0}^{\prime\prime}}{4\,\omega_{0}}-\frac{\omega_{0}^{\prime\,2}}{8\,\omega_{0}^{2}},\quad
\rho^-_{2}=-\frac{\omega_{0}^{\prime\prime}}{4\,\omega_{0}}+\frac{3\,\omega_{0}^{\prime\,2}}{8\,\omega_{0}^{2}},\\
\rho^+_{3}&=\frac{\omega_{0}^{\prime\prime\prime}}{12\,\omega_{0}}
-\frac{\omega_{0}^{\prime}\,\omega_{0}^{\prime\prime}}{8\,\omega_{0}^{2}}+\frac{\omega_{0}^{\prime\,3}}{16\,\omega_{0}^{3}},\\
\rho^-_{3}&=-\frac{\omega_{0}^{\prime\prime\prime}}{12\,\omega_{0}}
+\frac{3\,\omega_{0}^{\prime}\,\omega_{0}^{\prime\prime}}{8\,\omega_{0}^{2}}-\frac{5\,\omega_{0}^{\prime\,3}}{16\,\omega_{0}^{3}}, \\
\rho^+_{4}&=\frac{\omega_{0}^{\prime\prime\prime\prime}}{48\,\omega_{0}}
-\frac{\omega_{0}^{\prime}\,\omega_{0}^{\prime\prime\prime}}{24\,\omega_{0}^{2}}-\frac{\omega_{0}^{\prime\prime\,2}}{32\,\omega_{0}^{2}}
+\frac{3\,\omega_{0}^{\prime\,2}\omega_{0}^{\prime\prime}}{32\,\omega_{0}^{3}}-\frac{5\,\omega_{0}^{\prime\,4}}{128\,\omega_{0}^{4}},\\
\rho^-_{4}&=-\frac{\omega_{0}^{\prime\prime\prime\prime}}{48\,\omega_{0}}
+\frac{\omega_{0}^{\prime}\,\omega_{0}^{\prime\prime\prime}}{8\,\omega_{0}^{2}}+\frac{3\,\omega_{0}^{\prime\prime\,2}}{32\,\omega_{0}^{2}}
-\frac{15\,\omega_{0}^{\prime\,2}\omega_{0}^{\prime\prime}}{32\,\omega_{0}^{3}}+\frac{35\,\omega_{0}^{\prime\,4}}{128\,\omega_{0}^{4}}.
\end{align*}
Here $\omega_{0}^{\prime}$ is the group velocity of the carrier wave packet, $\omega_{0}^{\prime\prime}$ is the second-order dispersion coefficient, $\omega_{0}^{\prime\prime\prime}$ and $\omega_{0}^{\prime\prime\prime\prime}$ are high-order dispersion coefficients:
\begin{gather*}
\omega_{0}^{\prime}=\frac{c^2k_0}{\omega_0},\quad \omega_{0}^{\prime\prime}=\frac{c^2\alpha_1}{\omega_0^3},\quad
\omega_0^{\prime\prime\prime}=-\frac{3c^{4}k_0\alpha_{1}}{\omega_0^{5}},\\
\omega''''_{0}=\frac{3\,c^{4}\alpha_{1}}{\omega_0^{7}}\left(4c^{2}k_0^{2}-\alpha_{1}\right).
\end{gather*}

\section{Derivation of evolution equation for $u$}
The Lagrangian density for the nKG equation (\ref{nKG}) is
\begin{equation}\label{L}
L=\tfrac{1}{2}\phi_{t}^{2}-\tfrac{1}{2}c^{2}\phi_{x}^{2}-V(\phi).
\end{equation}
The wave field $\phi(x,t)$ given by ansatz (\ref{phianz}) is a function of the fundamental harmonic $\varphi$ and third harmonic $\phi_3$ that are considered as independent variables. Then the kinetic energy density in (\ref{L}) can be written as
\begin{equation*}
K=\tfrac{1}{2}\phi_{t}^{2} = \tfrac{1}{2}\varphi_{t}^{2}+\tfrac{1}{2}(\phi_3)_{t}^{2}+\varphi_{t}(\phi_3)_{t}.
\end{equation*}
The last (cross) term in $K$ disappears after averaging over the fast phase $k_0x_0-\omega_0t_0$. Therefore, it can be omitted in calculating the generalised momenta for the fields $\varphi$ and $\phi_3$:
\begin{equation*}
p=\frac{\partial L}{\partial\varphi_t}=\varphi_t,\quad
p_3=\frac{\partial L}{\partial(\phi_3)_t}=(\phi_3)_t.
\end{equation*}
Hamilton's principle \cite{Goldstein} formulated in a phase space formed by the fields $\varphi$, $\phi_3$ and their momenta $p$, $p_3$ requires the functional
\begin{equation}
S[\varphi,p,\phi_3,p_3]=\int\bigl(p\,\varphi_t+p_3(\phi_3)_t-\langle H\rangle\bigr)\,dx
\end{equation}
to keep a stationary value, so that $\delta S = 0$. Here $H$ is the Hamiltonian density given by formula~(\ref{H}) and $\langle\cdot\rangle$ means averaging over the fast phase. Taking variations of $S$ with respect to $p$ and $\varphi$, we come to a system of Hamilton's equations for these variables:
\begin{equation}\label{eqH1}
\varphi_{t}=\frac{\delta \mathcal{H}}{\delta p},\quad
p_{t}=-\frac{\delta\mathcal{H}}{\delta\varphi},
\end{equation}
where $\mathcal{H}=\int \langle H \rangle\, dx$ is the averaged Hamiltonian of the nKG equation.

Next we proceed to the equation for the complex symplectic coordinate $z$. By differentiating relation (\ref{z}) with respect to $t$ and taking into account Eqs.~(\ref{eqH1}), we get
\begin{equation*}
iz_{t}=\frac{1}{\sqrt{2}}\left(\sqrt{\widehat{\omega}}\,i\frac{\delta \mathcal{H}}{\delta p}
+\frac{1}{\sqrt{\widehat{\omega}}}\,\frac{\delta\mathcal{H}}{\delta\varphi}\right).
\end{equation*}
Calculating the functional derivatives
\begin{equation*}
\frac{\delta\mathcal{H}}{\delta p}=\frac{i}{\sqrt{2\widehat{\omega}}}
\left(\frac{\delta\mathcal{H}}{\delta z}-\frac{\delta\mathcal{H}}{\delta \overline{z}}\right),\quad
\frac{\delta\mathcal{H}}{\delta \varphi}=\sqrt{\frac{\widehat{\omega}}{2}}
\left(\frac{\delta\mathcal{H}}{\delta z}+\frac{\delta\mathcal{H}}{\delta \overline{z}}\right),
\end{equation*}
we come to the Hamilton equation for the function $z$:
\begin{equation}\label{eqH}
iz_{t}=\frac{\delta\mathcal{H}}{\delta \overline{z}}.
\end{equation}

Finally, the evolution equation (\ref{utau}) for the complex amplitude $u$ of symplectic coordinate $z$ is obtained from Hamilton's equation (\ref{eqH}) by substituting relation (\ref{z_trial}) and expanding the differential operator $\partial_t$ with the use of rule (\ref{dt}).

\section{Fifth-order NLSE for the non-canonical amplitude A}

Equation (\ref{NLSE5}) for the amplitude $u$ of the complex symplectic coordinate $z$ can be rewritten in terms of the complex amplitude $A$ of the first harmonic $\varphi$. To this end, we use relation (\ref{A_vs_u}) that expresses $A$ in terms of $u$ and differentiate it with respect to $\tau$. The terms with derivatives $u_\tau$ that appear in the right-hand side of the differentiated expression are calculated with the use of Eq.~(\ref{NLSE5}). After some algebraic transformations, we come to the following fifth-order NLSE for the amplitude $A$:
\pagebreak[0]
\begin{equation}\label{NLSE5_A}
\begin{split}
i\varepsilon^{2}&\bigl(A_{\tau}+\omega'_{0}\,A_{\chi}\bigr)
+\varepsilon^3\bigl(\tfrac{1}{2}\omega''_{0}A_{\chi\chi}+q^{(3)}|A|^{2}A\bigr)\\
+&\;i\varepsilon^{4}\bigl(-\tfrac{1}{6}\omega'''_{0}A_{\chi\chi\chi}+q_{1}^{(4)}|A|^2A_{\chi}+q_{2}^{(4)}A^{2}\overline{A}_{\chi}\bigr)\\
+&\;\varepsilon^{5}\bigl(-\tfrac{1}{24}\omega_{0}^{\prime\prime\prime\prime}A_{\chi\chi\chi\chi}
+q_{1}^{(5)}|A|^{4}A+q_{2}^{(5)}|A|^2A_{\chi\chi} \\
&\;\;+q_{3}^{(5)}A^{2}\overline{A}_{\chi\chi}+q_{4}^{(5)}A|A_{\chi}|^2+q_{5}^{(5)}\overline{A}A_{\chi}^{2}\bigr)=0,
\end{split}
\end{equation}
where
\begin{align*}
q^{(3)}&=\frac{\omega_{0}}{2}Q^{(3)}=-\frac{3\alpha_{3}}{8\omega_{0}},\\
q_{1}^{(4)}&=2\,q_{2}^{(4)}=\frac{\omega_{0}}{2}Q^{(4)}=-\frac{3\alpha_{3}c^{2}k_0}{8\omega_{0}^{2}}, \\
q_{1}^{(5)}&=\frac{\omega_{0}^{2}}{4}Q_{1}^{(5)}+\frac{\gamma}{2}Q^{(3)}=
\frac{1}{16\omega_0}\Bigl(\frac{9\alpha_{3}^{2}}{8\omega_0^{2}}-\frac{3\alpha_{3}^{2}}{16\alpha_{1}}-5\alpha_{5}\Bigr),\\
q_{2}^{(5)}&=\frac{\omega_{0}}{2}Q_{2}^{(5)}=\frac{3\alpha_{3}c^{2}}{8\omega_{0}^{5}}(2c^{2}k_0^{2}-\alpha_{1}),\\
q_{3}^{(5)}&=\frac{\omega_{0}}{2}\bigl(Q_{3}^{(5)}-2\,\rho_{1}^{2}\,Q^{(3)}\bigr) +\frac{\omega_{0}^{\prime\prime}\gamma}{2\omega_0}
=\frac{3\alpha_{3}c^{4}k_0^{2}}{8\omega_{0}^{5}},\\
q_{4}^{(5)}&=\frac{\omega_{0}}{2}\bigl(Q_{4}^{(5)}+2\,(\rho_{1}^{2}-2{\rho_{2}})\,Q^{(3)}\!+2{\rho_{1}}\,Q^{(4)}\bigr)
+\frac{\omega_{0}^{\prime\prime}\gamma}{\omega_0} \\
&=\frac{3\alpha_{3}c^{2}}{8\omega_{0}^{5}}(4c^{2}k_0^{2}-\alpha_{1}),\quad \rho_{1}\equiv\rho^-_{1},\;\;\rho_{2}\equiv\rho^-_{2},\\
q_{5}^{(5)}&=\frac{\omega_{0}}{2}\bigl(Q_{5}^{(5)}+(\rho_{1}^{2}-2{\rho_{2}})\,Q^{(3)}\bigr)
+\frac{\omega_{0}^{\prime\prime}\gamma}{2\omega_0}=\tfrac{1}{2}q_{4}^{(5)}.
\end{align*}

In contrast to Eq.~(\ref{NLSE5}) for $u$, Eq.~(\ref{NLSE5_A}) for the non-canonical amplitude $A$ is a non-Hamiltonian PDE. In particular, it contains the non-Hamiltonian term $A^{2}\overline{A}_{\chi}$ with $q_2^{(4)}\ne0$ that is absent in the Hamiltonian equation (\ref{NLSE5}). This proves that the coordinate-momentum coupling introduced by formula (\ref{z}) is a pivotal step in deriving a high-order NLSE in Hamiltonian form.

The coefficients of the equation for $A$ are expressed in terms of the coefficients of the equation for $u$. Their explicit expressions fully coincide with the same coefficients of the high-order NLSE for the amplitude $A$ derived in Refs.~\cite{ND18,ND19} by the methods of multiple scales and averaged Lagrangian. This fact proves the full correspondence between the results obtained by three different approaches of classical mechanics in application to the theory of nonlinear wave modulation.

\end{document}